\documentclass[preprint,showpacs,preprintnumbers,amsmath,amssymb]{revtex4}

\usepackage{graphicx}
\usepackage{dcolumn}
\usepackage{bm}

\begin{document}

\preprint{APS/123-QED}

\title{Quantitative analysis  of ferroelectric domain imaging \\ with piezoresponse force microscopy}

\author{Tobias Jungk}
\author{\'{A}kos Hoffmann}
\author{Elisabeth Soergel}

\email{soergel@physik.uni-bonn.de}

\affiliation{Physical Institute, University of Bonn,
Wegelerstra\ss e 8, 53115 Bonn, Germany}

\date{\today}

\begin{abstract}
The contrast mechanism for ferroelectric domain imaging via
piezoresponse force microscopy (PFM) is investigated. A novel
analysis of PFM measurements is presented which takes into account
the background caused by the experimental setup. This allows, for
the first time, a quantitative, frequency independent analysis of
the domain contrast which is in good agreement with the expected
values for the piezoelectric deformation of the sample and satisfies
the generally required features of PFM imaging.
\end{abstract}


\maketitle


Domain engineering in ferroelectric crystals is of increasing
importance for quasi-phase-matched second-harmonic generation
\cite{Fej92}, nonlinear photonic crystals \cite{Bro00}, and
ultra-high density data storage devices \cite{Cho02}. Among the
techniques utilized for the visualization of ferroelectric domains
\cite{Soe05} piezoresponse (or piezoelectric) force microscopy has
become an established standard tool because of its non-destructive
imaging capability with high lateral resolution \cite{Alexe,Par05}.
This detection technique is based on the deformation of the sample
due to the converse piezoelectric effect. The piezoresponse (or
piezoelectric) force microscope (PFM) is a standard scanning force
microscope (SFM) operated in contact mode with an additional small
alternating voltage applied to the tip. In piezoelectric samples
this voltage causes thickness changes and therefore vibrations of
the surface which lead to oscillations of the cantilever that can be
read out with a lock-in amplifier. However, although widely used,
the contrast mechanism for domain detection with PFM is still under
discussion mainly because of inconsistencies of the measured data
that concern the following features:
\begin{list}{-}
\item Frequency dependence:
the domain contrast should be independent of the frequency of the
alternating voltage applied to the tip. This applies of course only
for frequencies far away from any intrinsic resonance frequencies of
the cantilever. As the mechanical resonances of bulk ferroelectric
crystals are very high, they are irrelevant for our considerations
\cite{Bur75}.
\item Vibration amplitude:
the vibration amplitude of a $+z$ and a $-z$ domain face must be
equal. Its value $\Delta t$ should be in agreement with the
theoretical prediction  $\Delta t = d \cdot U$ with $d$ being the
appropriate piezoelectric constant and $U$ the voltage applied to
the tip \cite{Lin03}.
\item Phase shift: a phase difference of 180$^{\circ}$ between the
piezoelectric response on a $+z$ and on a $-z$ domain face is
considered mandatory.
\item Cantilever stiffness:
the domain contrast should be independent of the stiffness of the
cantilever used.
\end{list}

However, frequency scans of the alternating voltage applied to the
tip are reported to show a complex spectrum, i.e.~the measured
domain contrast strongly depends on the frequency. The vibration
amplitude measured is not equal on differently orientated domains
and the reported values differ by orders of magnitude. The phase
difference of 180$^{\circ}$ is not generally obtained. Finally the
domain contrast in PFM measurements was observed to be affected by
the stiffness of the cantilever. See e.g.
Refs.~\cite{Kol95,Eng98,Lab00,Lab01,Hong02,Har02,Har03,Har04,Scr05a,Agr05}.

Indeed, because of these basic inconsistencies with the features
listed above alternative origins for the domain contrast in PFM
measurements have been discussed. For the same experimental setup
the term ''dynamic-contact electrostatic force microscopy'' (DC-EFM)
was introduced and domain contrast was explained by specific
electrical properties of the $+z$ and the $-z$ domain faces
\cite{Hon98}. Differences in the work functions were also proposed
for causing the domain contrast \cite{Shv02}. To achieve a deeper
insight the electrostatic and the electromechanical contributions of
the tip-surface junction were calculated taking into account the
field and potential distributions as well as the indentation force
of the tip \cite{Kal02}.

Even though numerous approaches for an understanding of the PFM
contrast mechanism of ferroelectric domains have been reported, a
full (quantitative) analysis is still lacking. Although there is no
doubt that PFM imaging is sensitive to ferroelectric domains, the
opposite situation (a contrast in PFM imaging unambiguously proving
the existence of ferroelectric domains) is not yet established
because of the above mentioned inconsistencies. A more
detailed understanding of the PFM detection method is therefore
needed.

In this letter we present a novel analysis of the data acquired with
PFM. This allows for the first time a clear understanding of the
contribution of the converse piezoelectric effect which is found to
fully satisfy the features listed above.

For the investigations, we used a conventional experimental setup
with a commercial scanning force microscope (SMENA, NT-MDT),
modified to allow application of voltages to the tip. We utilized
four different cantilevers C$_{1}$-C$_{4}$ with Pt/Ir-coated tips
(Micromasch) of lengths $100 - 130\,\rm \mu m$, resonance
frequencies $160 - 290$\,kHz, and stiffness
$\rm C_{1}$:~$k$~=~5.3\,N/m,
$\rm C_{2}$:~$k$~=~9.8\,N/m,
$\rm C_{3}$:~$k$~=~11.4\,N/m and
$\rm C_{4}$:~$k$~=~26.4\,N/m.
For PFM operation we applied an alternating voltage (amplitude:
10\,V$_{\rm pp}$) to the tip and detected the resulting oscillation
of the cantilever with a lock-in amplifier (SRS 830), the phase
being set to $0^{\circ}$, the time constant to 3\,ms. We
simultaneously recorded the in-phase ($\theta = 0^{\circ}$) and the
orthogonal ($\theta = 90^{\circ}$) output, $\theta$ denoting the
phasing with respect to the alternating voltage applied to the tip.
In the following these output signals of the lock-in amplifier will
be named PFM signals, $p$ and $n$ being the PFM signal on a $+z$ and
a $-z$ domain face respectively. The sample was a periodically
poled, $z$-cut, congruently melting lithium niobate crystal ($\rm
8\times 10 \times 0.5$\,mm$^3$) with a period length of 8\,$\mu$m.

The experimental procedure was as follows: we firstly recorded a PFM
image of the sample in order to subsequently position the tip
accurately on a $+z$ or a $-z$ domain face. We then measured the
frequency dependence of the amplitude of the cantilever oscillations
by scanning the alternating voltage applied to the tip from 10\,kHz
to 100\,kHz. The scan duration was about 10\,minutes. The graphs in
this letter are averages over three separate frequency scans taken
at different positions on the sample surface.

\begin{figure}
\includegraphics{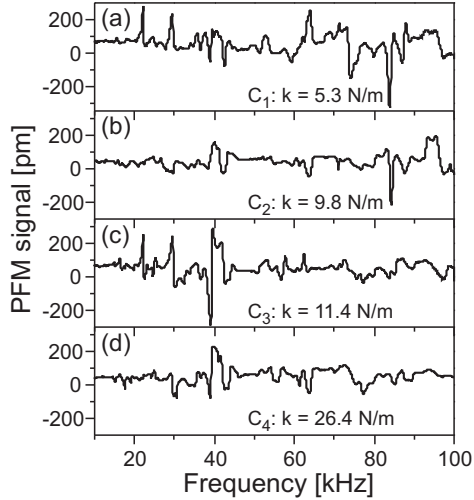}
\caption{\label{fig:Jungk1} Frequency dependence of the in-phase PFM
signal on a $+z$ domain face of PPLN for four different cantilevers
C$_{1}$-C$_{4}$, k: spring constant.}
\end{figure}

Figure \ref{fig:Jungk1} shows frequency scans of the in-phase PFM
signal on a $+z$ domain face for the four different cantilevers
used. The frequency spectra look apparently random, although some
specific features recur (for example at $\sim$22 and at
$\sim$29\,kHz for C$_1$ and C$_3$ and at $\sim 84$\,kHz for C$_1$
and C$_2$). The PFM signal reaches values of more than 250\,pm
whereas only 75\,pm are predicted for the surface vibration due to
the converse piezoelectric effect in $z$-cut $\rm LiNbO_3$. Moreover
at some specific frequencies, no PFM signal is measured and even
negative values are obtained. The PFM signal of the orthogonal
output of the lock-in amplifier shows a similar behavior, however,
with completely different spectra.

Frequency spectra similar to the ones shown in Fig.~\ref{fig:Jungk1}
have already been reported \cite{Lab00,Har04,Scr05a}. For their
explanation the excitation of resonant modes of the cantilever was
proposed, tip and sample surface being in contact with each other.
\cite{Lab00}. We observed, however, that frequency scans with no
sample in the vicinity of the tip result in similar spectra,
admittedly with a smaller amplitude. If the tip is in contact with
the sample, the frequency spectrum can be affected e.g.~by changing
the coupling conditions between the tip and the SFM head. We
therefore wrapped the silicon chip (to which the tip is attached)
with conductive scotch tape. This led to an altered spectrum with
much larger amplitudes. These results indicate a complex mechanical
resonance behavior of the whole setup comprising the sample, the tip
with cantilever and the SFM head. From the spectra shown in
Fig.~\ref{fig:Jungk1} and the findings described above, it is
obvious that only a small part of the PFM signal on $\rm LiNbO_3$
can be attributed to the ferroelectric properties of the sample. The
PFM signal is completely dominated by a complex background signal.

\begin{figure}
\includegraphics{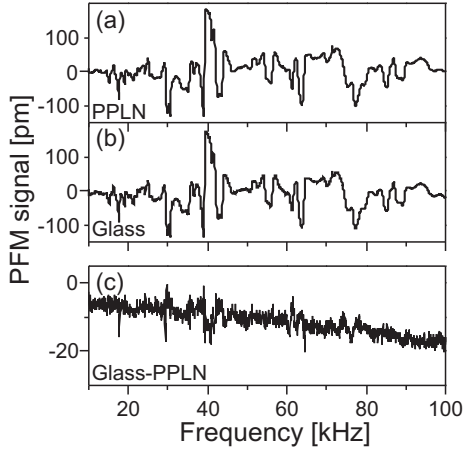}
\caption{\label{fig:Jungk2}
Frequency dependence of (a) the in-phase background PFM signal on a
PPLN surface $( b= \frac {p + n}{2})$ , (b) the in-phase PFM signal
on a glass surface and (c) the difference between these two graphs.
The measurements were performed with the cantilever  C$_{4}$.}
\end{figure}

To determine this background signal, we averaged over the $+z$ and
$-z$ domain faces: $(p+n)/2$, therefore eliminating the
contributions of the ferroelectric properties of the sample to the
PFM signal (Fig.~\ref{fig:Jungk2}(a)). To prove this statement we
performed reference measurements with the same cantilever on a
standard glass microscope slide (Fig.~\ref{fig:Jungk2}(b)). The
difference between these two frequency spectra is shown to be
extremely small (Fig.~\ref{fig:Jungk2}(c) the vertical scale being
expanded by a factor of ten). The slight decrease towards higher
frequencies might be due to a drift of the experimental setup during
the scan time of 10\,minutes. The graphs clearly show a
reproducible, frequency dependent PFM signal independent of the kind
of sample used. In the following this PFM signal will be denoted as
the background PFM signal $b=(p+n)/2$.

\begin{figure}
\includegraphics{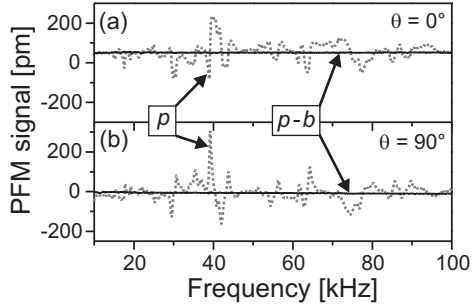}
\caption{\label{fig:Jungk3} Frequency dependence of the PFM signal
on a $+z$ domain face of PPLN: (a) in-phase and (b) orthogonal
output. The dotted gray curves $p$ show the measured PFM signal, the
black curves $p-b$ the calculated, background-corrected PFM signal.
The measurements were performed with the cantilever C$_{4}$.}
\end{figure}

In order to extract the contributions of the ferroelectric
properties of the $\rm LiNbO_3$ sample from the PFM signal we
subtracted the background PFM signal from the measured data. The
result is shown in Fig.~\ref{fig:Jungk3} on a $+z$ domain face for
the in-phase (a) and the orthogonal (b) output of the lock-in
amplifier. The background-corrected curves ($p-b$, black lines) are
plotted together with the measured PFM signals ($p$, gray lines). As
can be clearly seen, the part of the PFM signal causing the domain
contrast appears only in phase with the applied voltage with a
constant amplitude.

\begin{figure}
\includegraphics{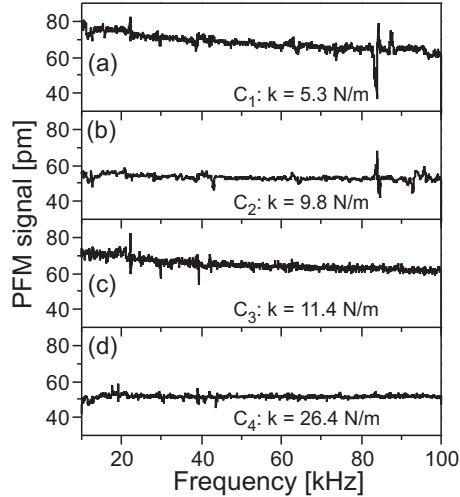}
\caption{\label{fig:Jungk4}Frequency dependence of the in-phase,
background-corrected PFM signal on a $+z$ domain face of PPLN for
four different cantilevers C$_{1}$-C$_{4}$, k: spring constant.}
\end{figure}

For an interpretation of the background-corrected PFM signal we
performed a quantitative analysis of the measurements.
Figure~\ref{fig:Jungk4} shows the frequency spectra of the
background-corrected in-phase PFM signals for the four cantilevers,
the vertical scale being expanded by a factor of ten with respect to
Fig.~\ref{fig:Jungk3}. All cantilevers show a frequency independent
spectrum, the averaged values are C$_1$:~62.1\,pm, C$_2$:~58.8\,pm,
C$_3$:~70.5\,pm, and C$_4$:~51.8\,pm. This has to be compared with
the theoretically expected value for the converse piezoelectric
effect of $\Delta t = \frac{\varepsilon_{333}}{C_{333}} \cdot U =
75$\,pm with $\varepsilon_{333} = 1.785$\,C/m$^2$ and $C_{333}=2.357
\times 10^{11}$\,N/m$^2$ being the appropriate piezoelectric and
stiffness tensor elements respectively \cite{Jaz02}.

Although the background-corrected PFM signals are of the right
magnitude, they are all too small however by 5 - 30\,\% as compared
to the theoretically expected value
\footnote{This has to be compared to published values that vary from
20\,pm for KTP ($\rm d_{33}\sim 20\,pm/V$) \cite{Eng98} to 30\,nm
for GASH \cite{Kol95} ($\rm d_{33}\sim 2\,pm/V$), both with 10\,V
applied to the tip.}.
A possible explanation lies in the mechanical constrictions of the
deformation. The electrical field at the tip, causing the
piezoelectric deformation, spatially decays extremely fast due to
the small radius of curvature of the tip ($\sim 30$\,nm)
\cite{Kol95}. As a consequence, the thickness changes of the crystal
occur in a volume comparable to the tip size. Because of its
stiffness, the crystal cannot fully follow the required deformation
which could be the cause for measuring too small values. Using
larger tips should result in higher values for the piezoelectric
deformation
\footnote{Note that although at the very tip, the electric field
might be as high as 10$^7$\,V/m (with 10\,V applied to the tip),
this has no influence on the theoretically expected piezoelectric
thickness change which is determined only by the applied voltage
\cite{Lin03}.}.

An important point here is that the PFM signal was found to be
independent of the stiffness of the cantilever. Because we always
operate the PFM at the same set-point of the feedback circuit
(i.e.~the same bending of the cantilever), the graphs in
Fig.~\ref{fig:Jungk4} this already indicate that the indentation of
the tip has no influence on the PFM signal. We confirmed this
statement by using a stiff cantilever and varying the set-point,
thus changing the indentation force by two orders of magnitude. The
observed frequency spectra remained mainly unchanged. Note that a
too strong indentation can trigger a local switching of the
polarization of the material \cite{Alp01}.

With the results described above, the contrast mechanism in PFM
imaging of ferroelectric domains can be fully explained through the
thickness change of the sample due to the converse piezoelectric
effect, taking into account the background PFM signal as determined
above.

\begin{figure}
\includegraphics{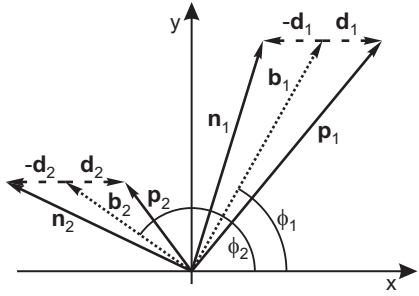}
\caption{\label{fig:Jungk5} Vector diagram for the domain contrast
in PFM measurements exemplified for two different frequencies
$\omega_{1}$ and $\omega_{2}$. The $x$-axis denotes the in-phase
output ($\theta = 0^{\circ}$) and the $y$-axis the orthogonal output
($\theta = 90^{\circ}$) of the lock-in amplifier. In the graph ${\bf
b}_{1}$, ${\bf b}_{2}$ denote the background PFM signals and
$\phi_{1}$, $\phi_{2}$ their phases, ${\bf p}_{1}$, ${\bf p}_{2}$
and ${\bf n}_{1}$, ${\bf n}_{2}$ are the measured PFM signals on a
$+z$ and on a $-z$ face respectively and $2{\bf d}_1 =  2 {\bf d}_2$
is the domain contrast. The background PFM signal rotates randomly
with frequency, changing its phase and amplitude which strongly
affects the measured PFM signals although the domain contrast is
constant.}
\end{figure}

To summarize the situation, a vector diagram illustrates the case
for two different frequencies $\omega_1$ and $\omega_2$ of the
alternating voltage applied to the tip (Fig.~\ref{fig:Jungk5}). At a
certain frequency $\omega_1$, a background PFM signal ${\bf b}_1$ is
present. The ferroelectric domains contribute ${\bf d}_1$ for the
$+z$ face and ${\bf -d}_1$ for the $-z$ face to the PFM signal, both
of same amplitude with a 180$^{\circ}$ phase shift between. This
results in the measurement of ${\bf p}_1 = {\bf b}_1 + {\bf d}_1$
for the $+z$ face and ${\bf n}_1={\bf b}_1 - {\bf d}_1$ for the $-z$
face. It is important to note that the phasing between ${\bf p}_1$
and ${\bf p}_2$ is not $180^{\circ}$, their amplitudes are unequal
and larger than the expected value. The same considerations apply of
course for any other frequency $\omega_2$. It is obvious from
Fig.~\ref{fig:Jungk5} that although the domain contrast is the same
($2{\bf d}_1 =  2 {\bf d}_2$) the PFM signals measured at different
frequencies differ with respect to amplitude and phase.

In conclusion we have presented a novel analysis of the detection
mechanism of ferroelectric domains with piezoresponse force
microscopy. Taking into account the background PFM signal caused by
the whole experimental setup, basic inconsistencies in PFM
measurements concerning frequency dependence, amplitude, phasing and
stiffness of the cantilever could be removed. Thus the origin of the
domain contrast on PPLN could be explained solely via the converse
piezoelectric effect, satisfying the generally required features of
PFM imaging. The experimental data were found to be in good
agreement with the theoretically expected values. Performing a
quantitative analysis of the PFM signal it can thus be determined
whether an observed contrast in PFM imaging can be attributed to the
converse piezoelectric effect of the sample, therefore unambiguously
proving the existence of domains in ferroelectric materials.


\begin{acknowledgments}
We thank R.W. Eason for stimulating discussions.
Financial support of the DFG research unit 557 and of the Deutsche
Telekom AG is gratefully acknowleged.
\end{acknowledgments}


\end{document}